\documentclass[twocolumn,secnumarabic,amssymb, nobibnotes, aps, prl, reprint]{revtex4-2}
\usepackage[english]{babel}
\usepackage{bm}
\usepackage{xcolor}
\usepackage{amsmath}
\usepackage{graphicx}
\usepackage{hyperref}

\newcommand{\GW}{$G_0W_0$}
\newcommand{\VBM}{VBM$@\Gamma$}
\newcommand{\sCBm}{sCBm}
\newcommand{\CBmG}{CBm$@\Gamma$}
\newcommand{\CBmX}{CBm$@\Xi$}

\begin{document}
\title{Spatially indirect excitons in black and blue phosphorene double layers}
%\title{Optical absorption and charge separation in black/blue phosphorene heterostructures}

\author{Michele Re Fiorentin}%
\email{michele.refiorentin@iit.it}
\affiliation{Center for Sustainable Future Technologies, Istituto Italiano di Tecnologia, via Livorno 60, 10144, Torino, Italy}
\author{Giancarlo Cicero}%
\affiliation{Dipartimento di Scienza Applicata e Tecnologia, Politecnico di Torino, corso Duca degli Abruzzi 24, 10129, Torino, Italy}
\author{Maurizia Palummo}
\affiliation{Dipartimento di Fisica and INFN, Universit\`{a} di Roma ``Tor Vergata'', via della Ricerca Scientifica 1, 00133, Roma, Italy}
\date{\today}%

\begin{abstract}
Monolayer black and blue phosphorenes possess electronic and optical properties that result in unique features when the two materials are stacked. We devise a low-strain van-der-Waals double layer and investigate its properties with {\it ab initio} many-body perturbation theory techniques. A type-II band alignment and optical absorption in the visible range are found. The study demonstrates that spatially indirect excitons with full charge separation can be obtained between two layers with the same elemental composition but different crystalline structure, proving the system interesting for further studies where dipolar excitons are important and for future opto-electronic applications.
\end{abstract}

\maketitle
%\tableofcontents

\noindent\textbf{Cite as:} Phys. Rev. Materials \textbf{4}, 074009 (2020), \href{https://doi.org/10.1103/PhysRevMaterials.4.074009}{doi.org/10.1103/PhysRevMaterials.4.074009}

\section{Introduction}
Over the past decades, a wide range of two-dimensional (2D) materials have been synthesized in the wake of the crucial discovery of graphene \cite{geim}. 2D materials such as hexagonal boron nitride \cite{hBN}, metal carbides and nitrides \cite{MXenes1,MXenes_review}, transition metal dichalcogenides (TMDs) \cite{TMDs_review} and single-element monolayers such as silicene \cite{silicene} and germanene \cite{germanene}, have been under extensive experimental and theoretical scrutiny because of their remarkable physical, electronic and optical properties \cite{2D_1,2D_2,Bernardinano16,Palummo2019}. 
Recently, two mono elemental 2D materials based on phosphorus, namely black-phosphorene ({black-P}) \cite{blackP_synthesis1,blackP_synthesis2,blackP_synthesis3} and blue-phosphorene (blue-P) \cite{blueP_synthesis1,blueP_synthesis2} have joined the list.

Black-P has a puckered honeycomb structure and presents a direct electronic gap. In the literature, the value of the electronic bandgap for the monolayer (ML) form ranges from 1.6 to 2.4~eV, depending on the experimental or theoretical technique used \cite{Liu2014,Tran2014,Peeters2014,Rudenko2014,Tran2015,blackP_gap_exp,Li2017}.
For freestanding ML, a value of 2.4 eV has been recently predicted by accurate quantum Monte-Carlo simulations \cite{Frank19}.
A similar spread, from 1.2 to 1.7 eV \cite{Tran2014,Peeters2014,blackP_GW1,Li2017} affects the measured and theoretically predicted optical gap values. Nevertheless, a  large exciton binding energy is always found, due to the electronic quantum confinement and to the reduced dielectric screening typical of low-dimensional materials \cite{low_dimension1}. Moreover, Raman and IR/VIS optical spectra of black-P show a distinctive anisotropy in the material plane \cite{blackP_nat_comm,blackP_raman1}. 

Blue-P is an allotrope less studied than black-P, that displays a hexagonal lattice. 
The ML form has been demonstrated to be stable at room temperature \cite{blueP_theo_prediction} and presents an indirect electronic gap whose predicted values range from 2.98 to 3.56~eV \cite{blueP_GW1,Peng2016,Han2017,Shu2019,blueP_GW2}. Blue-P shows a very large exciton binding energy ($\sim1$~eV) and isotropic optical absorption in the material plane \cite{Peng2016,blueP_GW1}. 

As often proposed in the case of many 2D materials, the electronic and optical features of black and blue phosphorene can be exploited jointly, through the realization of vertical heterostructures \cite{heterojunctions}, in which different layers are stacked by exploiting long-range, weak van-der-Waals (vdW) interactions. 
Indeed, heterostructures can be of great interest for opto-electronic applications, depending on the resulting alignment of the two monolayer electronic bands. A type-II alignment is desired for applications in photocatalysis \cite{photocatalysis1,photocatalysis2} or  photovoltaics  \cite{photovoltaics,bernardi_palummo}, since it favors the separation of the photo excited electron and hole pairs. 
The physical models currently reported to achieve charge separation are based on interfaces between two different materials showing a type II heterojunction \cite{hetero_review,heterojunctions}, between homojunctions with different doping ($p-n$ junctions) \cite{homojunctions} or between organic/inorganic structures \cite{organic} with optimal molecular orbital alignments at the interface with the solid surface. Here, we demonstrate that effective electron/hole separation can be achieved in a bidimensional bilayer system with the same composition (elemental phosphorous) but different crystalline structure, without the need for dopants. Moreover, we find that in such layered systems a small applied strain is enough to make the system change from indirect to direct absorber. The observed mechanism is general and can be in principle extended to other 2D materials existing in different crystalline structures: this opens up the possibility of observing the phenomena in a wide range of absorption energies (thus applications), depending on the band gap of the starting monolayers.

In addition, the search for spatially indirect excitons in 2D vdW heterostructures is a topic of recent interest also from the fundamental point of view \cite{Fogler2014}. Thanks to their long lifetimes and large binding energy, spatially indirect excitons can cool below the quantum degeneracy temperature, which is expected to be higher than what found in traditional GaAs heterostructures \cite{GaAs_QW1}.
Beyond TMD hetero bilayers \cite{NL_Calman20}, double layers of phosphorene seem to be particularly appealing in this context, due to the predicted high-temperature electron-hole anisotropic superfluidity \cite{PRB_Pouya,PRB_Berman}. 

Double layers based on phosphorene allotropes have been recently proposed \cite{black_blue_1,black_blue_2}, but the suggested structures show a high stretching of the monolayers, in comparison to their respective equilibrium geometries. This implies a reduced structure stability and a strong perturbation of the electronic properties which, especially for low-dimensional materials, are extremely sensitive to strain \cite{2D_strain,2D_strain_palummo}. Moreover, these previous studies are limited to density functional theory (DFT) only and thus lack of an accurate discussion of the role of many-body effects that are crucial to correctly describe the electronic and excitonic properties of nanomaterials.

\section{Methods}
In this work we set up a black-P/blue-P van-der-Waals double layer in which the two layers undergo a strain lower than 1\% with respect to their equilibrium lattice parameters and investigate its electronic and optical properties by performing many-body perturbation theory simulations on top of DFT ones. 
We carry out non-self-consistent $GW$ (\GW{}) calculations \cite{hedin,GW1} to correct the DFT electronic band energies and precisely determine the bandgap. The obtained quasi-particle energies are then employed in the solution of the Bethe-Salpeter equation (BSE) \cite{bethe_salpeter,BSE1}, yielding optical properties with real predictive value. 
We discuss the obtained optical spectrum and the spatial localization of the first excitonic state, finding that a partial charge separation on the two layers can be achieved. Finally, we show how the application of mild uniaxial strain leads to a reshuffling of the lowest conduction band states and results in the appearance of an excitonic state at lower energy,  with small dipole strength but with a complete charge separation.
\begin{figure}
\centering
\includegraphics[width=1\linewidth]{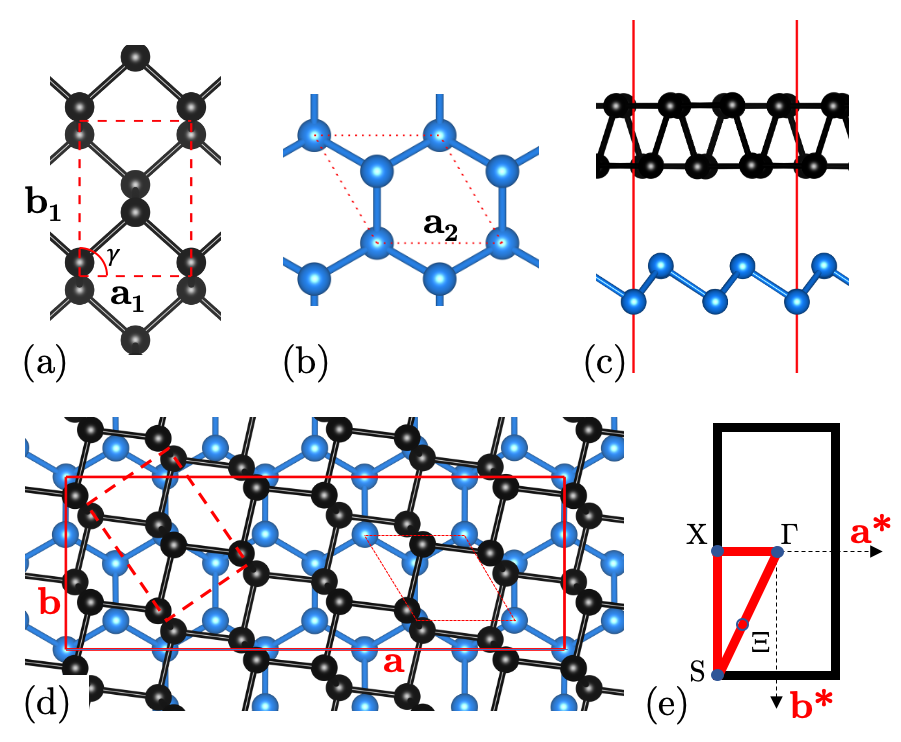}
\caption{(a), (b) Optimized geometries of single layer black and blue phosphorenes respectively. The primitive cells are marked with dashed and dotted red lines. (c) Side view of the heterostructure unit cell. (d) Top view of the optimized geometry of the black/blue phosphorene heterostructure. The unit cell is marked by the solid red line. Dashed (dotted) thin red lines represent the single layer black-P (blue-P) primitive cells, for comparison. (e) Path (red line) along the Brillouin zone with marked critical points and the special $k$ point $\Xi$. }
\label{fig:geometry}
\end{figure}
The double layer unit cell is identified by means of the CellMatch \cite{cellmatch} code starting from the primitive cells of black-P and blue-P monolayers.
The geometry optimizations of single phosphorene layers  and of the heterostructure are performed within the plane-wave approach to DFT of the Quantum ESPRESSO package \cite{qe1,qe2} (cf. Supplementary Material). 
The vdW interaction, responsible for the stability of the double layer, is reproduced with the semi-empirical Grimme-D3 method \cite{grimme-D3}, whose predictions on black-P geometry have been shown to be in remarkably good agreement with the results obtained with quantum Monte Carlo techniques \cite{QMC_grimme}. 
In fig.~\ref{fig:geometry}(c) and (d) we report the side and top views of the optimized double layer. For comparison, in fig.~\ref{fig:geometry}(a) and (b) we report the primitive cells of single layer black-P and blue-P, respectively. The largest stretching in the double layer is found along the $\mathbf{b}_1$ direction of black-P, whose corresponding lattice parameter changes from 4.62~\AA{} in the optimized monolayer to 4.66~\AA{} in the double layer, corresponding to a ${\sim{}0.8}$\% strain. The angle $\gamma$ between $\mathbf{a}_1$ and $\mathbf{b}_1$ is increased from 90$^\circ$ to $90.9^\circ$ in the double layer. The other parameters undergo smaller variations. We conclude that the double layer configuration we propose shows very little strain with respect to the monolayer equilibrium parameters, thus ensuring that the electronic properties are not artificially skewed. The obtained double layer is computed to be stable, yielding a cohesion energy $E_\textup{coh}=-14\,\mbox{meV/\AA}^2$ with respect to the separate layers.

\section{Results and discussion}
\begin{figure}[h]
\centering
\includegraphics[width=\linewidth]{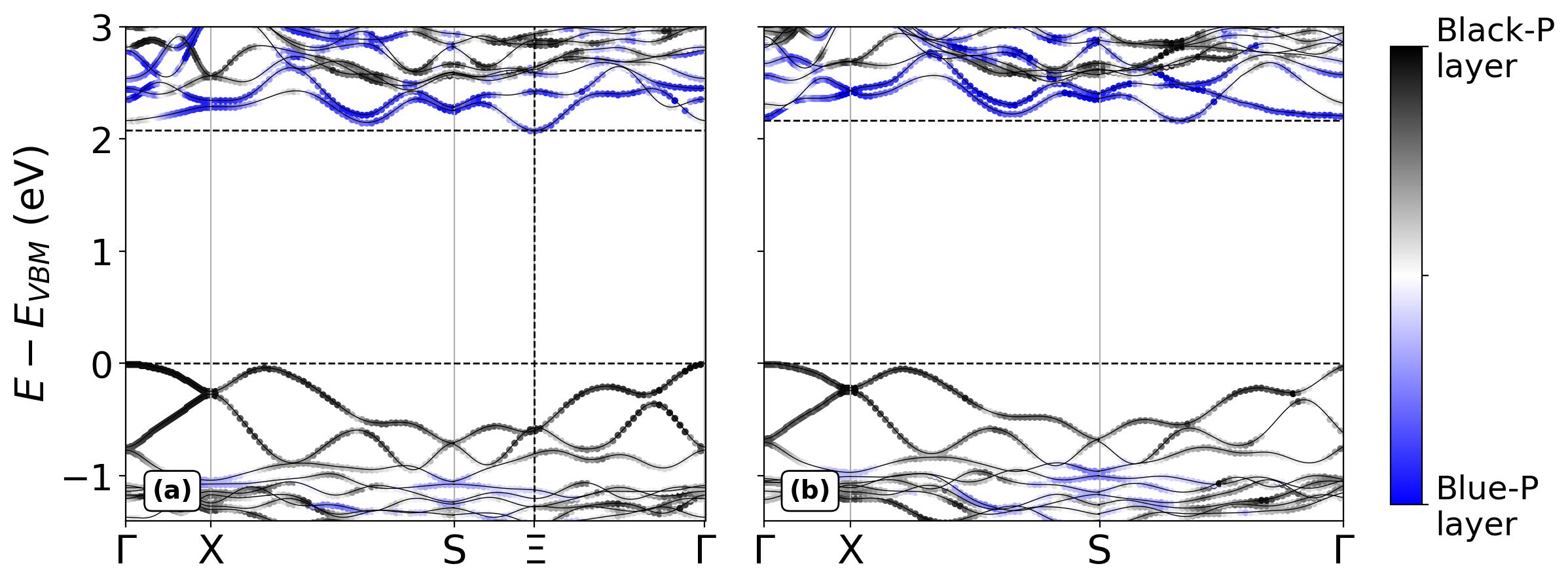}
\caption{\GW{} corrected band structure and $k$-resolved DOS of the unstrained (a) and strained heterostructure (b). The band coloring along the $k$-path marks the difference between the DOS contributions from the atomic orbitals belonging to the two layers, ranging from blue (contribution from blue-P atoms only), to white (equal contribution from both layers), to black (contribution from black-P atoms only).}
\label{fig:bands}
\end{figure}

The electronic properties of the optimized double layer are studied within the \GW{} approximation by means of the YAMBO code \cite{yambo1,yambo2} (cf. Supplementary Material). 
In fig.~\ref{fig:bands}(a) we report the \GW{} corrected band structure along the path shown in fig.~\ref{fig:geometry}(e). For each band and $k$-point, we project the electronic wavefunction on the orbitals of the P atoms and identify which is the contribution to the density of states (DOS) stemming from the atoms of one each layer. We color the bands along the $k$-path according to the difference between the DOS contributions from the two layers. This way, we are able to evaluate if the electronic states at various $\mathbf{k}$ are localized in one layer or in the other or in both.  From fig.~\ref{fig:bands}(a) we notice first that the blue-P/black-P double layer shows an indirect electronic bandgap of 2.08~eV between the valence band at the $\Gamma$ point and the conduction band at an intermediate $k$-point $\Xi$ between $S$ and $\Gamma$. It is, then, interesting to see how the largest contribution to the electronic state at the valence band maximum (\VBM) at the $\Gamma$ point (\VBM$@\Gamma$) comes from orbitals of phosphorus atoms belonging to the black-P layer, while the state at the conduction band minimum (CBm), at the $\Xi$ point (\CBmX) is entirely due to blue-P orbitals. At the $\Gamma$ point, the direct gap is 2.17~eV and the wavefunction of the first unoccupied state (\CBmG) receives a significant contribution from both layers, as marked by a lighter shade in fig.~\ref{fig:bands}(a), and the electronic band shows little dispersion along the $\overline{\Gamma X}$ direction. On the contrary, the state at the next-to-bottom of the conduction band at the $\Gamma$ point (CBm+1$@\Gamma$) is dominated by the contribution from blue-P atoms and the energy strongly disperses along $\overline{\Gamma X}$. We mention in passing that DFT calculations largely underestimate the indirect bandgap by about 1.24~eV.

\begin{figure}[h]
\centering
\includegraphics[width=\linewidth]{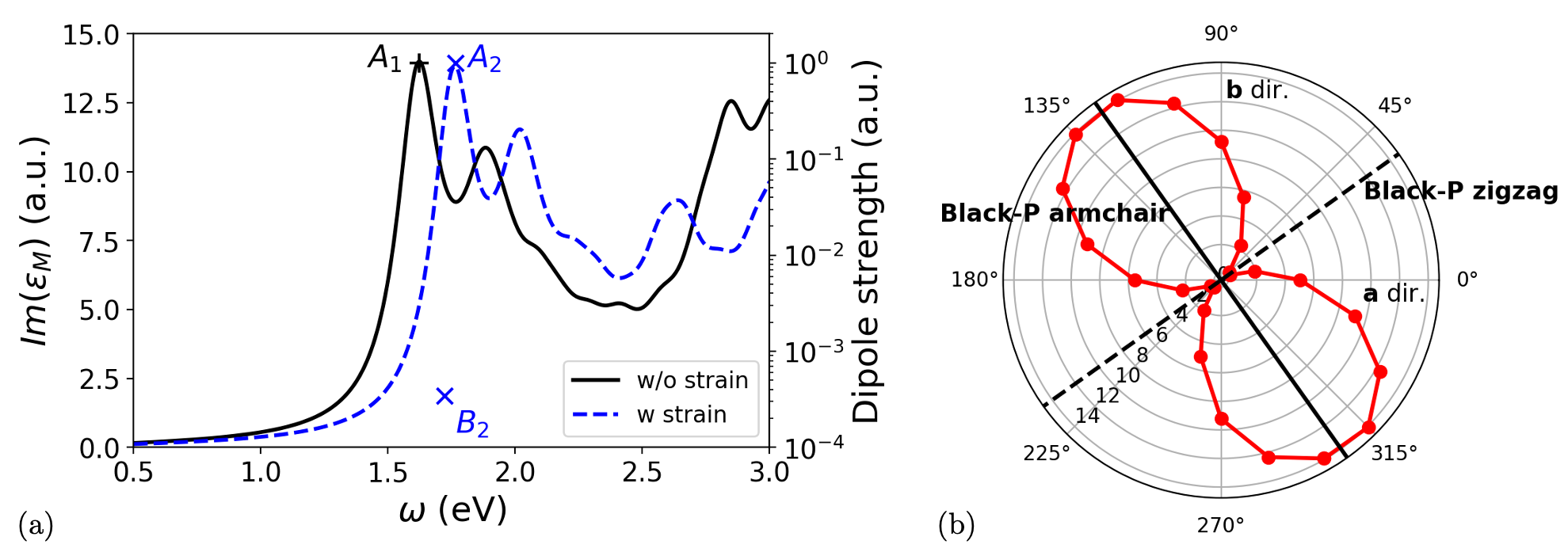}
\caption{(a) Optical absorption spectrum, $\mbox{Im}(\varepsilon_M)$ in arbitrary units, obtained from the solution of the BSE for the unstrained (solid black line) and strained (dashed blue line) heterostructure. The dipole strength scale (arbitrary units) refers to the marked exciton $A_1$, $A_2$ and $B_2$. (b) Polar plot of the absorption peak amplitude vs. polarization angle of incoming, in-plane polarized photons. The $0^\circ$ and $90^\circ$ directions are aligned to the heterostructure $\mathbf{a}$ and $\mathbf{b}$ vectors, respectively.}
\label{fig:BSE}
\end{figure}

\begin{figure}[h]
\centering
\includegraphics[width=\linewidth]{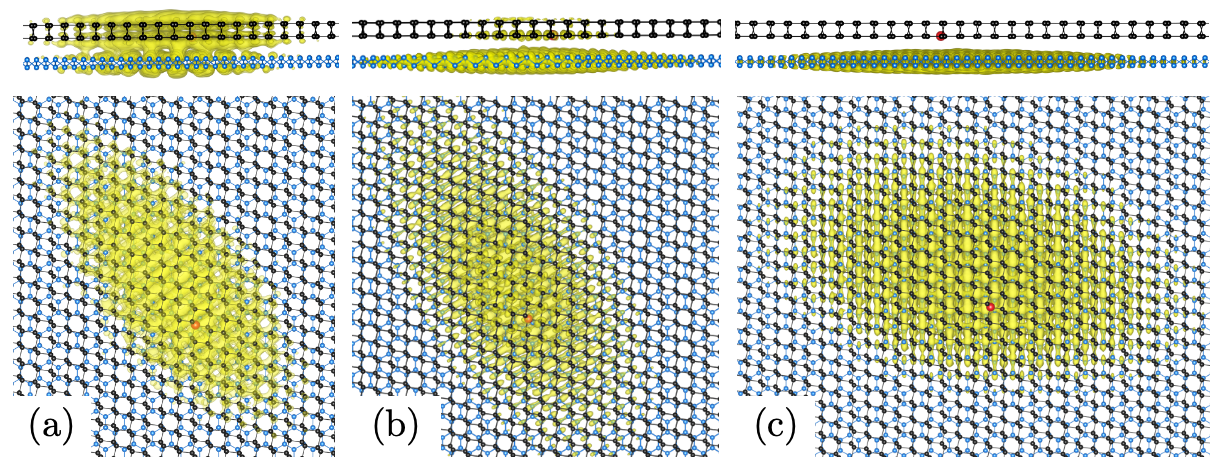}
\caption{Square modulus of the exciton wavefunction as a function of the electron position with fixed hole position. Side (upper panels) and top (lower panels) view of $A_1$ (a), $B_1$ (b) and $B_2$ (c). The hole is marked by the red sphere.}
\label{fig:exc}
\end{figure}

The optical absorption spectrum is computed by solving the BSE with the YAMBO code (cf. Supplementary Material) and is reported in fig.~\ref{fig:BSE}(a).%
We consider an incoming photon linearly polarized in the heterostructure plane along the black-P armchair direction. The heterostructure shows a marked optical anisotropy imposed by the black-P layer, as shown in fig.~\ref{fig:BSE}(b). Here it is possible to notice that the absorption is largest and smallest along the black-P armchair and zig-zag directions, respectively. From the black, solid line spectrum in fig.~\ref{fig:BSE}(a) we deduce the presence of a sharp excitonic peak, $A_1$, for an incident photon energy $E_{A_1}=1.62$~eV. This bright exciton state also corresponds to the lowest energy solution of the BSE and is dominated by the transition between the VBM and the CBm+1 around the $\Gamma$~point. From the electronic gap at the $\Gamma$ point we obtain the $A_1$ binding energy to be 0.55~eV. This is in line with the large values typical of 2D materials, though slightly smaller \cite{blackP_gap_exp,blackP_GW1,blueP_GW1} due to the interaction between the two layers resulting in an increased electronic screening with respect to the monolayers. We can gain information on the localization of $A_1$ by plotting the square modulus of its wavefunction, $\left|\Psi_{A_1}(R_h,r_e)\right|^2$, as a function of the electron position $r_e$, while the hole is kept fixed at $R_h$. We choose to position the hole in the regions of space in which the square modulus of the \VBM{} wavefunction is largest, i.e. in the vicinity of phosphorus atoms on the black-P layer. The obtained isosurface is shown in fig.~\ref{fig:exc}(a).
Here, the hole is marked by the red sphere on the black-P layer and the isosurface highlights that the paired electron is delocalized on both layers, having a non-null probability of being found on blue-P. However, as already noticed while discussing the orbital composition of \CBmG, the contribution of black-P orbitals is significant, resulting in a low net charge separation and the marked spatial anisotropy typical of black-P \cite{blackP_gap_exp}. Although $A_1$ is the lowest-energy exciton state with null center of mass momentum $\mathbf{Q}$, considering non-null momenta reveals the presence of a dark exciton, $B_1$, with $\mathbf{Q}\simeq\Xi$ and a lower energy, $E_{B_1}=1.57$~eV. $B_1$ receives a large contribution from the transition between \VBM{} and \CBmX{} and its localization is reported in fig.~\ref{fig:exc}(b). As could be anticipated by the study of the orbital character of \CBmX{} in fig.~\ref{fig:bands}(a), $\left|\Psi_{B_1}(R_h,r_e)\right|^2$, with fixed hole on black-P, is largely localized on the blue-P layer, resulting in a high charge separation probability. On the other hand, $\mathbf{Q}\neq0$ implies that $B_1$ cannot be directly excited by light, but necessarily requires phonon scattering.

The band structure of single layer phosphorenes can be effectively altered by the application of strain \cite{blackP_strain1,blackP_strain2,blueP_strain1}. Aiming at generating a direct bandgap heterostructure, we apply a mild uniaxial strain to the system, increasing by 2\% the $\mathbf{a}$ vector and study the resulting modification to the electronic properties, fig.~\ref{fig:bands}(b). The strain along the $\mathbf{a}$ direction greatly affects the electronic bands with a large dispersion along the $\overline{\Gamma X}$ direction in the Brillouin zone, while states composed of more localized orbitals, which disperse less, show a much smaller variation in energy. For this reason, the electronic band structure remains mostly unchanged around the top valence bands, while it is dramatically modified within the conduction band at the $\Gamma$ point. Here, the \CBmG+1 state, with large dispersion, leaps under the \CBmG{} state, which shows a much more reduced dispersion, and the electronic bandgap becomes direct at the $\Gamma$ point at 2.17~eV. This way, the conduction band bottom of the strained structure (\sCBm) is now entirely given by blue-P orbitals, fig.~\ref{fig:bands}(b). The optical absorption spectrum of the strained heterostructure is shown by the dashed blue line in fig.~\ref{fig:BSE}(a). As in the unstrained case, the spectrum shows a marked excitonic peak, $A_2$, at $E_{A_2}=1.76$~eV, dominated by the transition at the $\Gamma$ point from VBM to \sCBm+1. As mentioned above, the orbital character of VBM does not change from the unstrained case, while \sCBm+1 is identical to \CBmG. Therefore, the localization of $A_2$ is the same as for $A_1$ in fig.~\ref{fig:exc}(a). In the strained case, the BSE shows another solution, $B_2$, at $E_{B_2}=1.72$~eV, around 40~meV below $A_1$. This lower-energy solution is characterized by the transition between VBM and \sCBm{} at the $\Gamma$ point, so that the excited electron state is completely dominated by the contributions from blue-P orbitals, cf. fig.~\ref{fig:bands}(b). Hence, given the localization of VBM on the black-P layer, we expect that exciton $B_2$ enhances the charge separation by confining the hole on black-P and the electron on blue-P. This is confirmed by $\left|\Psi_{B_2}(R_h,r_e)\right|^2$ as plotted in fig.~\ref{fig:exc}(c). Here, the red spot marking the hole is fixed on the black-P layer and consequently the conditional probability of finding the electron is non-negligible only on the facing blue-P layer. Moreover, the dominant blue-P character is highlighted by the less marked anisotropy of the exciton wavefunction \cite{blueP_GW1}. However, as shown by the cross in fig.~\ref{fig:BSE}(a), the dipole strength of $B_2$ exciton is about four orders of magnitude smaller than $A_2$, hence, though not truly dark, $B_2$ appears very difficult to directly excite through photon absorption in comparison to $A_2$.

The analysis of the exciton spectra of the unstrained and strained heterostructures shows that, in both cases, the strongest absorption peaks $A_1$ and $A_2$, though resulting in delocalized exction wavefunctions, exhibit a moderate charge separation between the two layers. Nevertheless, the BSE presents an additional solution at a lower energy both in the unstrained ($B_1$ with $\mathbf{Q}=\Xi$) and in the strained case ($B_2$ at the $\Gamma$ point with $\mathbf{Q}=0$). It is then possible to envisage a phonon-mediated mechanism able to yield $B_1$ and $B_2$ from the optical excitation of $A_1$ and $A_2$, respectively \cite{phon_relax1,phon_relax2}. In the unstrained case, with the creation of the $A_1$ exciton, the electron can be found on the blue-P layer, opposite to the hole. The presence in the phonon spectrum of blue-P of LO and TO phonons, showing a marked Einstein-like dispersion with  $\omega_\textup{LO/TO}\simeq50$~meV \cite{blueP_phonon_spectrum}, suggests the possibility of a scattering process with optical phonons carrying $\mathbf{Q}\simeq\Xi$ and $\omega=E_{A_1}-E_{B_1}\simeq50$~meV, resulting in the relaxation of the system from $A_1$ to $B_1$. Being an energy minimum and featuring a net charge separation between the two layers, exciton $B_1$ will then radiatively recombine with a largely suppressed rate \cite{interlayer_phonon}. Similarly, in the strained case, the system can efficiently relax from the optically excited $A_2$ state to the spatially separated $B_2$ exciton through the exchange of $\mathbf{Q}=0$ phonons. The creation of interlayer excitons via this polarization-to-population transfer \cite{polariz_to_popul1,polariz_to_popul2} has been recently shown to be extremely efficient even at low temperatures in the case of vdW heterostructures of 2D TMDs \cite{interlayer_phonon}.

\section{Conclusions}
In summary, we have studied a low-strain vdW double layer consisting of ML black and blue phosphorenes that shows a 2.08~eV indirect electronic bandgap at the \GW{} level. The solution of the BSE resulted in a bright excitonic peak at 1.62~eV with a partial charge separation between the two layers. We have shown how the band structure can be effectively modified through the application of mild uniaxial strain, resulting in a reshuffling of the band states at the bottom of the conduction band at the $\Gamma$ point and in the consequent 2.17~eV direct electronic bandgap. The bright excitonic peak shifts to 1.76~eV, keeping a partial charge separation. In both cases, we have found that the BSE presents an exciton solution at lower energy: with finite momentum in the unstrained double layer and at the $\Gamma$ point in the strained case. Though with null or small dipole strength, and thus impossible or hard to directly excite, we have suggested that the polarization-to-population transfer  can take place in our systems, obtaining an efficient population of the lower-energy excitonic states characterized by a net charge separation between the two layers. 
Our results prove that effective electron/hole separation can be achieved in a homo-elemental 2D bilayer system, thus providing an intriguing starting point for further theoretical studies and opening up new promising prospects in a wide range of optoelectronic fields.

We acknowledge the CINECA award under Iscra-B and Iscra-C initiatives, for the availability of high performance computing resources, as well as the computational facilities and support provided by HPC@POLITO (http://www.hpc.polito.it). M.P. acknowledges fnancial support from the EU MSCA HORIZON2020 Project DiSeTCom (GA 823728).

%\bibliographystyle{science}
%\bibliography{black_blueP_manuscript}

\end{document}